\documentclass{article} 
\usepackage{nips11submit_e}
\usepackage{times}
\usepackage{algorithm}
\usepackage{algorithmic}
\usepackage{amssymb}
\usepackage{amsmath}
\usepackage{xspace}
\usepackage{multirow}
\usepackage{graphicx}
\usepackage{enumitem}
\setitemize{itemindent=-0.01cm,leftmargin=13pt,itemsep=2pt}

\title{Image Retrieval based on Bag-of-Words model}

\author{
Jialu Liu \\
Department of Computer Science\\
University of Illinois at Urbana-Champaign\\
Urbana, IL, USA \\
\texttt{jliu64@illinois.edu} \\
}

\begin{document}

\maketitle

\begin{abstract}
This article gives a survey for bag-of-words (BoW) or bag-of-features model in image retrieval system.
In recent years, large-scale image retrieval shows significant potential in both industry applications and research problems.
As local descriptors like SIFT demonstrate great discriminative power in solving vision problems like object recognition, image classification and annotation, more and more
state-of-the-art large scale image retrieval systems are trying to rely on them.
A common way to achieve this is first quantizing local descriptors into visual words, and then applying scalable textual indexing
and retrieval schemes. We call this model as bag-of-words or bag-of-features model.
The goal of this survey is to give an overview of this model and introduce different strategies when building the system based on this model.

\end{abstract}

\section{Introduction}
In recent years, large-scale image retrieval shows significant potential in both industry applications
\footnote{Google Image: http://www.google.com/imghp} and research problems.
Depending on the query formats, image retrieval is roughly divided into two
categories: text-based approaches and content-based methods.
The text-based approaches associate keywords with each stored image and user usually only needs to type in keywords to search for images.
No obvious differences exist between this text-based approach and traditional text retrieval systems.
However, researchers find that sometimes it may be difficult to describe image content with a small set of keywords, which motivates
Content-based image retrieval (CBIR). It tries to grasp the semantic similarity between the content between images directly.

The problem of searching images according to their semantic content is very challenging since there exist many factors affecting the performance like
resolution, illumination variations and occluded objects.
While global features are known to be limited in face of these difficulties, which describe a picture in a holistic way (eg. one histogram represents a fingerprint for the whole image),
local descriptors like Scale Invariant Feature Transform (SIFT) \cite{sift} provide a way to describe several salient patches around key points within the images,
demonstrating great discriminative power. The cardinality of the set of local descriptor vectors depends on the detected key points in the picture,
resulting a huge number to cope with for the large-scale retrieval system.
The most popular approach today, initially proposed in \cite{videogoogle},
relies on a bag-of-words (BoW) or bag-of-features model. The idea is to
first quantize local descriptors into ``visual words'' and represent each image as a vector of words like one document.
Then we can mimic text-retrieval systems, applying scalable indexing and fast search on this vector space.
It is noted that the BoW approach, although is simple and directly borrowed from text retrieval community,
has shown excellent performance not only for CBIR task but also for other vision tasks like object recognition, image classification and annotation \cite{surveyobjectrecognition, sampling}.

\section{Bag-of-words Representation}
In this section, we introduce the bag-of-words (BoW) model in a more detailed way and point out several main procedures which will be discussed in the later sections.

\begin{figure}
  \center
  \includegraphics[width=5in]{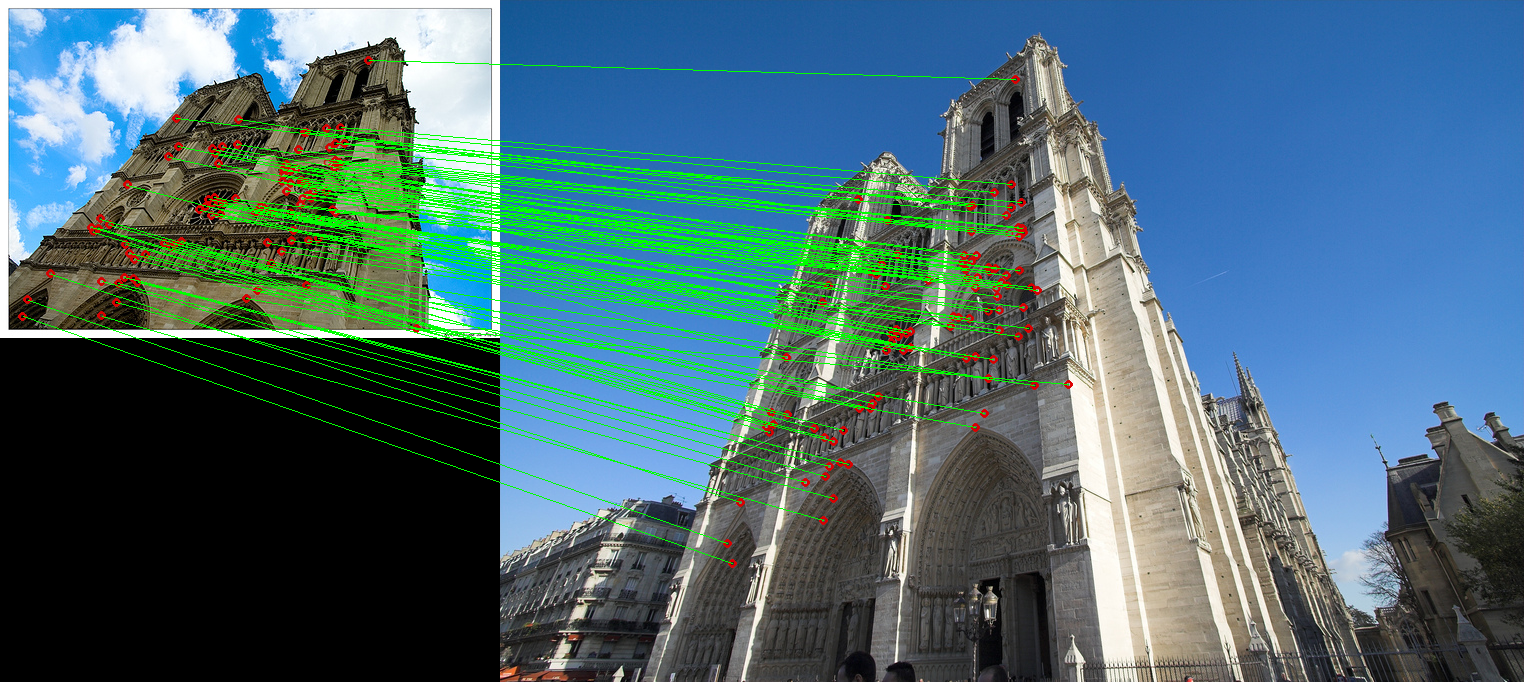}\\
  \caption{SIFT feature matching based on DoG keypoints \footnote{Obtained from web: http://acmechimera.blogspot.com}}\label{fig:match}
\end{figure}

State-of-art large-scale image retrieval systems have relied on adopting the BoW model initially proposed in \cite{videogoogle}.
BoW model in CBIR is mainly designed for the local descriptors of images which describe regions around the key points detected in the images.
Different from global features describing a picture in a holistic way, one image can have a bunch of salient patches around key points. And local descriptor
like 128-dimensional SIFT is demonstrated to be a good way to represent the characteristics of these patches. Figure \ref{fig:match} gives a intuitive way to do SIFT
match.
But once we extract such local descriptors for each image, the total number of them might be huge.
And searching nearest neighbors for each local descriptor in the query image becomes too time consuming.
Therefore, BoW is proposed as a way to solve this problem by quantizing descriptors into ``visual words'', which decreases the descriptors' amount dramatically.
In this way, each image can be viewed as a long and sparse vector of words and therefore we can mimic text-retrieval systems, applying scalable indexing and fast search on this vector space.

In BoW model, the following several procedures must be taken into consideration.
\begin{itemize}
  \item Choose methods to detect key points and describe local patches around key points. This step can be viewed as preprocess of BoW model and it is mostly related to computer vision in this paper.
        How to detect and describe key points should be critical to the retrieval performance since all the following procedures are relied on this ``data generation'' step.
  \item Quantize local descriptors into ``visual words''. We call this procedure as vocabulary generation and local descriptor quantization.
        The fundamental difference of BoW model in text and image retrieval is that
        text words are sampled naturally according to language context; visual words are the outcomes of quantization. It is artificially generated and infers statistical information which has less semantic sense.
        So how to generate such a vocabulary remains an interesting question and different methods are proposed toward this direction.
  \item Indexing and search. Once we obtain the vocabulary and quantization method, we can mimic text-retrieval systems, applying scalable indexing and fast search on this vector space. Inverted file has been proved
        to be an efficient and common way to achieve this goal.
\end{itemize}

Although BoW model has shown excellent performance in image retrieval task, it still suffers from some problems.
One major concern is that BoW model could not support spatial properties of local descriptors of images.
Several papers were then published to improve this framework from different angles so that spatial information is taken into account.

Another disadvantage is that key points are extracted from the grey level images and local descriptors do not contain any color information \cite{towards}.
Moreover, they only grasp the local information, losing the overall distribution of visual information. The performance could be enhanced if we can combine global features and local descriptors together.

Finally, some papers claim that strategies common in text retrieval community like query expansion could be adopted to enrich the BoW model.

\section{Key point Detector and Local Descriptor}
Both key point detection and local description can be considered as preprocess of BoW model.
Typically they together form the basis of BoW model. Because this section is closely related to computer vision and there already exist outstanding survey 
\cite{surveyobjectrecognition}, we just briefly introduce some related works.

\subsection{Key point Detector}
Key point detector samples a sparse set of locally stable points (and their support regions) \cite{towards}.
These sampled key points are expected to be invariant to geometric and photometric changes.
Different detectors, nevertheless, emphasize different aspects of invariance, resulting in
key points of varying properties and sampled sizes.

Here we introduce seven popular key point detectors introduced in \cite{towards, surveyobjectrecognition}. They are:
\begin{itemize}
  \item Laplacian of Gaussian (LoG)
  \item Difference of Gaussian (DoG)
  \item Harris Laplace
  \item Hessian Laplace
  \item Harris Affine
  \item Hessian Affine
  \item Maximally Stable Extremal Regions (MSER)
\end{itemize}

In LoG, the scale-space representation is built by successive smoothing of high
resolution image with Gaussian based kernels of different
sizes.  A feature point is then detected if a local 3D extremum is present and if its absolute value is higher than
a threshold. The LoG detector is circularly symmetric and
it detects blob-like structures. In DoG, the input image is
successively smoothed with a Gaussian kernel and sampled.
The DoG representation is obtained by subtracting two successive smoothed images. 
Thus, all the DoG levels are constructed by combined smoothing and subsampling. The
DoG is an approximate but more efficient version of LoG.
The Harris Laplace detector responds to corner-like regions.
It uses a scale-adopted Harris function to localize points in
scale-space, and then selects the points for which the Laplacian of Gaussian attains a maximum over a scale.
Key points of Hessian Laplace are
points which reach the local maxima of Hessian determinant in space and fall into the local maxima of Laplacian of
Gaussian in a scale. Harris Affine, which is derived from Harris-Laplace, estimates the
affine neighborhood by the affine adaptation based on the
second moment matrix, while Hessian Affine is achieved after
the affine adaptation procedure based on Hessian Laplace.
MSER is a watershed-like detector based on intensity value.
The obtained regions are of arbitrary shape and they are defined by all the border pixels enclosing a region,
where all the intensity values within the region are consistently lower or higher with respect to the
surrounding.

\subsection{Local Descriptor}
Local descriptors describe the key points' neighborhood which is already identified by the detectors.
They are also expected to own some invariance properties. Here invariance means that the descriptors should be robust
against various image variations such as affine distortions, scale changes, illumination changes or compression artifacts \cite{surveyobjectrecognition}.
It is obvious that the descriptors' performance strongly depends on the power of the key point detectors.

Here we introduce three popular local detectors. They are:
\begin{itemize}
  \item Scale Invariant Feature Transform (SIFT)
  \item Speeded Up Robust Features (SURF)
  \item Locally Binary Patterns (LBP)
\end{itemize}

\begin{figure}
  \center
  \includegraphics[width=5in]{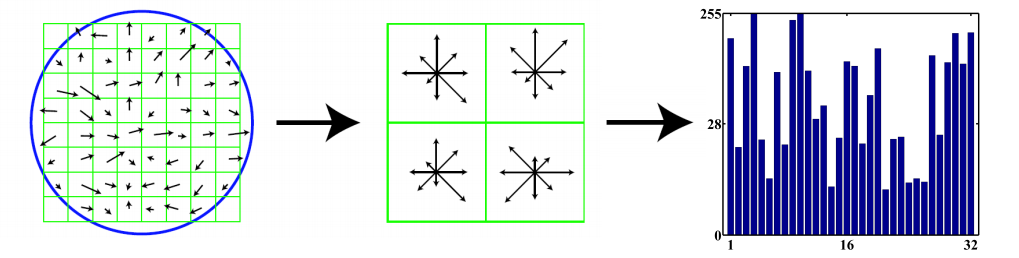}\\
  \caption{Illustration of the SIFT descriptor calculation partially taken from
\cite{distinctive}. Note that only a 32 dimensional histogram obtained from a $2 \times 2$ grid
is depicted for a better facility of illustration.}\label{fig:sift}
\end{figure}

In SIFT\footnote{SIFT originally bundles DoG key point detector with a proper descriptor called SIFT-key. In this paper, for convenience, SIFT refers to its descriptor.}\cite{sift},
The circular region around the key-point
is divided into $4 \times 4$ not overlapping patches and the histogram gradient orientations within these patches are calculated. Histogram smoothing is
done in order to avoid sudden changes of orientation and the bin size is
reduced to 8 bins in order to limit the descriptor's size. This results into
a $4 \times 4 \times 8 = 128$ dimensional feature vector for each key point. Figure \ref{fig:sift}
illustrates this procedure for a $2 \times 2$ window. Recently studies have shown that
SIFT is one of the best descriptors for key points.

SURF \cite{surf} is partly inspired by the SIFT descriptor, which is also a combination of detector and descriptor.
The standard version of SURF is several times faster than SIFT and claimed by its authors to be more robust against different image transformations than SIFT.
SURF is based on sums of 2D Haar wavelet responses and makes an efficient use of integral images.
It uses an integer approximation to the determinant of Hessian blob detector, which can be computed extremely quickly with an integral image (3 integer operations).
For descriptor, it uses the sum of the Haar wavelet components around the point of interest. Again, these can be computed with the aid of the integral image.

LBP is a very simple texture descriptor approach
initially proposed by \cite{lbp}. It is based on a very simple binary
coding of thresholded intensity values through comparing each pixel with its neighbors.
Then histogram of these binary codes is computed, normalized and concatenated.
LBP is invariant to monotonic gray value transformations but it is not inherently rotational invariant.
And partial scale invariance of the descriptors can be reached in combination
with scale invariant detectors.

Note that these local descriptors are all computed based on grey level images and do not contain any color information \cite{towards}.

\section{Vocabulary Generation and Local Descriptor Quantization}
To quantize local descriptors into visual words, we must first generate visual vocabulary.
Like document in text retrieval, each image in the ``corpus'' is represented as a sparse vector of ``term'' (visual word) occurrences
through local descriptor quantization, and search then proceeds by calculating the similarity between the query vector and each ``document'' vector.
Different from text retrieval where text words are sampled naturally according to language context, visual words are artificially generated which has less semantic sense.

\subsection{Vocabulary Generation}
A visual vocabulary is often generated by scalable clustering methods which view local descriptors as raw input.
By clustering, we treat each cluster as a unique visual word from the vocabulary. It is often the case that
centroids of these clusters are stored so that later each image including the query can be easily quantized into visual words.
It is worth noticing that generating vocabulary is done off-line and may be time consuming since the amount of local descriptors could be huge.

\subsubsection{Methods}
Early systems \cite{videogoogle} used a flat
$k$-means clustering that was effective but difficult to scale
to large vocabularies. More recent work has used approximate nearest neighbor search method \cite{object} or cluster hierarchies \cite{hkm} 
to greatly increase the visual-word vocabulary size.

\textbf{Approximate $k$-means}\\
Approximate $k$-means (AKM) extends flat $k$-means by adopting approximate nearest neighbor search method \cite{object}.  
Specifically, it uses randomized $k$-d trees to accelerate the process of finding nearest centroid.

Usually in a $k$-d tree, each node splits the dataset
using the dimension with the highest variance for all the
data points falling into that node and the value to split on
is found by taking the median value along that dimension
(although the mean can also be used). In the randomized
version, the splitting dimension is chosen at random from
among a set of the dimensions with highest variance and the
split value is randomly chosen using a point close to the median. The conjunction of these trees creates an overlapping
partition of the feature space and helps to mitigate quantization effects, where features which fall close to a partition
boundary are assigned to an incorrect nearest neighbor.
This robustness is especially important in high-dimensions,
where due to the “curse of dimensionality”, points will
be more likely to lie close to a boundary.

The algorithmic complexity of a single k-means iteration
is then reduced from $O(NK)$ to $O(N \log K)$, where N is
the number of descriptors being clustered from.

\textbf{Hierarchical $k$-means}\\
In \cite{hkm}, hierarchical $k$-means (HKM) is proposed
to generate a ``vocabulary tree''. On the first level of the tree,
all local descriptors are clustered into a small number (e.g., $K$ = 10) of cluster centers.
On the next level, $k$-means (with $K = 10$ again) is applied within each of the
partitions independently. So the final result is $K^n$ clusters at the $n_{th}$ level.
A new data point is assigned
by descending the tree. Instead of assigning each data point
to the single leaf node at the bottom of the tree, the points can additionally be assigned to some internal nodes which
their path from root to leaf passes through. This can help
mitigate the effects of quantization error, for cases when
the data point lies close to the Voronoi region boundary for
each cluster center.

It is important to note that traditional flat $k$-means minimizes the total sum of euclidian distance between the
local descriptors and their assigned closest cluster centers, whereas the hierarchical
tree minimizes it only locally at each node and
this does not in general result in a minimization of the total
sum of distance.

The algorithmic complexity of a single k-means iteration in HKM is also $O(N \log K)$ where the base of the logarithm is the
cluster number of each node in the vocabulary tree.

\subsubsection{Vocabulary Size}
While the size of text vocabulary is predefined by the corpus, the size of visual vocabulary is obtained by clustering methods.
Therefore, the chosen size of the vocabulary becomes tricky and interesting.
A small vocabulary may lack the discriminative
power since two local descriptors may be assigned into the same
cluster even if they are not similar to each other. A large
vocabulary, on the other hand, is less generalizable, less forgiving to noises, and incurs extra processing overhead \cite{towards}.
Additionally, distinct datasets may also prefer different size of vocabulary since the vocabularies themselves are not similar due to the topics of datasets.
Actually, the vocabulary size varies a lot as mentioned in different papers, from 1K to 1M.
But it is suggested that the broader dataset covers, the larger the vocabulary should be.
And constrained by the memory size and scale of computing power, large vocabulary could be impractical.

\subsection{Local Descriptor Quantization}
Local descriptor quantization is the process of assigning one local descriptor to one or multiple visual words. 
To assign a descriptor to visual word(s), the common way is to search for nearest neighbors among the vocabulary obtained in the vocabulary generation step. 

\subsubsection{Methods} \label{NNS}
Corresponding to the methods for generating vocabulary, there are also a few candidates for nearest neighbor search besides brute-force approach.
Additionally, hashing like methods try to achieve sublinear complexity against this task.

\textbf{Approximate Nearest Neighbor search}\\
Approximate nearest neighbor search is the quantization method adopted in AKM. It uses randomized $k$-d trees to accelerate the process of finding nearest centroid.
It reduces the complexity to $O(\log K)$ where $K$ is the total amount of words (centroids) in the vocabulary.

\textbf{Hierarchical Nearest Neighbor search}\\
Hierarchical nearest neighbor search is the quantization method adopted in HKM. 
It uses a ``vocabulary tree'' to accelerate nearest neighbor search and the time complexity is $O(\log K)$ where the base of the logarithm is the
cluster number of each node in the vocabulary tree.

\textbf{Hashing Like Nearest Neighbor search}\\
Hashing like nearest neighbor search is a family of methods for deriving low-dimensional discrete representations of the data with sublinear time complexity.

Among these method, locality-sensitive hashing (LSH) \cite{similarityhashing, similarityestimation}, offers probabilistic guarantees of retrieving items within some constant value times
the optimal similarity. The basic idea is to compute randomized hash functions that guarantee a high probability of collision for similar examples. 

Semantic hashing \cite{semantichashing} seeks compact binary codes of data-points so that the
Hamming distance between codewords correlates with semantic similarity.

Spectral hashing \cite{spectralhashing} approach is motivated by the idea that a good encoding scheme should minimize the sum of
Hamming distances between pairs of code strings weighted by the value of a Gaussian kernel between the corresponding feature vectors.

\subsubsection{Weight Schemes}
Term weighting is known to have critical impact to text retrieval. 
Whether such impact extends to visual keywords remains an interesting question.

\textbf{TF-IDF}\\
The standard weighting is known as ``term frequency-inverse document frequency'' (\emph{tf-idf}) \cite{improving, videogoogle}.
It is assumed that visual words appearing in more images should have lower weight when computing the similarity.
Some papers also adopted \emph{tf} directly \cite{beyond}.
It is reported in \cite{towards} that the impact of \emph{idf} is
sensitive to vocabulary size because a
frequent visual word may be split into several rare
words when increasing the vocabulary size. Thus
the \emph{idf} weight of a certain visual word is not stable at all.

\textbf{Binary}\\
Binary weighting indicates that the presence and absence of a visual word with values 1
and 0 respectively. It is a straightforward way of compacting a BoW vector since it discards
the exact number of occurrences of a given visual word in the image \cite{packing}.
In \cite{towards}, the experiments show that \emph{tf}  outperforms binary by
a large margin only when the vocabulary size is small. This
is due to the fact that, with a larger vocabulary size, the
count of most visual keywords is either 0 or 1 and thus \emph{tf}
features are similar with binary features.

\textbf{Soft}\\
In \cite{towards}, soft weighting scheme is proposed considering the distances from key points
to centroids, by selecting the top $k$ nearest visual words.
The closer a local descriptor is situated to a visual word, the higher its significance for that specific visual word.
The authors state that their approach outperforms the above two strategies across different vocabulary sizes.

\subsection{Open Source Implementations}
To our knowledge, there are two open source implementations of scalable $k$-means and approximate nearest neighbor search.

\textbf{FASTCLUSTER} library\footnote{http://www.robots.ox.ac.uk/~vgg/software/fastcluster/} is a python/C++ library for performing fast, distributed (using MPI)
AKM for very large datasets as described in \cite{object}.

\textbf{FLANN} \footnote{http://www.cs.ubc.ca/~mariusm/index.php/FLANN/FLANN} is a library for performing fast approximate nearest neighbor searches in high dimensional spaces. It contains a collection of algorithms we found to work best for nearest neighbor search and a system for automatically choosing the best algorithm and optimum parameters depending on the dataset \cite{ann}.
It is written in C++ and contains bindings for the following languages: C, MATLAB and Python.

\section{Indexing and Search}
The common way to deal with the retrieval is to rely on the inverted index where each 
visual word has a index indicating the images containing it together with weights.
The scores for each image are then accumulated so that they are identical to explicitly
computing the similarity \cite{object}.

There also exist other strategies to do the search besides inverted index like accelerated nearest neighbor search.

\subsection{Inverted Index}
In the worst case, the computational complexity of querying the index is linear in the
corpus size, but in practice it is close to linear in the number
of images that match a given query, generally a major
saving. For sparse queries, this can result in a substantial
speedup, as only images which contain words present in
the query need to be examined.

With large corpora of images, memory usage becomes
a major concern. And when main memory is exhausted,
the engine can be switched to use an inverted index flattened to disk, which caches the data for the most frequently
requested words \cite{object}.

\subsection{Scoring function}
\textbf{Hamming Distance}\\
For binary weighting scheme, as each image is represented as a binary code vector, 
we can directly compute the hamming distance between them.

\textbf{Cosine Similarity}\\
Cosine similarity is a measure of similarity between two vectors by measuring the cosine of the angle between them.

\textbf{$\mathbf{L_2}$ Distance}\\
The $L_2$ distance is also known as the Euclidean distance. 

\textbf{Hamming Embedding}\\
Hamming embedding \cite{improving} provides binary
signatures that refine visual words. It can be stored in the index together with each image ID.
Hamming embedding incorporates a similarity
measure for descriptors assigned to the same visual word with the assumption that 
local descriptors partitioned to the same visual word still differentiate from each other.
It is claimed to work well when the vocabulary is not big. However, this strategy results in both computational and spatial overhead.

\subsection{Nearest Neighbor Search}\label{NNS2}
If the vocabulary is not too large, we can view each image as a unit and directly adopt nearest neighbor search
methods which are discussed in \ref{NNS}.

\section{Additional Strategies}
To further improve the performance, several works are proposed from various angles considering problems BoW model suffers from.

\subsection{Spacial Information} \label{Spacial}
One major concern regarding BoW model is that it does not support spatial properties of local descriptors of images just like that in text retrieval.
Several papers were published to improve this framework so that spatial information is taken into account.

Among these attempts, RANSAC-based image reranking achieved the state-of-the-art result in terms of retrieval accuracy.
However, these methods require reranking top image search results and accessing local descriptors of these images, which
increases the overhead to a large degree \cite{object, totalrecall}.

``Bundled features'' were then proposed in \cite{bundling} trying to encode spatial information in stable regions (MSER + SIFT) in inverted index.
But the comparison of order information must be incorporated into the scoring function, making distance
measure no longer an L2 or other common distance, and thus cannot be
further accelerated by nearest neighbor search methods mentioned in \ref{NNS} and \ref{NNS2}.

Spatial-bag-of-features \cite{spatial}  adapts the orderless BoW to a so-called spatial BoW by
changing the order of the histograms; the spatial histogram of each visual word is rearranged by starting from
the position with the maximum frequency. However, this rearrangement may not correspond to the true transformation (translation, rotation and scaling).

\subsection{Combination with Global Features}
We know that in BoW model, key points are extracted from the grey level images and local descriptors do not contain any color information \cite{towards}.
Additionally, BoW model only grasps the local information, losing the overall distribution of visual information. The performance could be enhanced if we can combine global features and local descriptors together.

In \cite{fire}, the authors evaluate empirically tuned linear combinations, a trained
logistic regression model, and support vector machines to fuse different global features with SIFT histogram.

\subsection{Query Expansion}
Query expansion is a common strategy in text retrieval, which can be directly adopted to enrich the BoW model in image retrieval.

In \cite{totalrecall} and \cite{improving}, several query expansion methods are mentioned to improve the search performance.
Here we mention two of them:

\textbf{Transitive closure expansion} (TCE) considers the tree of
images with the initial query being its root. The children of a node are images that reliably match with it.
TCE consists of a breadth-first scan of the tree, where
nodes are returned as results in the order they are visited. The number of expansions is limited to 20 to
avoid drift.

\textbf{Additive query expansion} (AQE). The interest points of
reliable results to the initial query are geometrically re-mapped to this image. The resulting set of points is
used to perform a second query. The returned images
are appended to results of the initial query. Only one
re-querying is performed.

To assure the effect of query expansion, it is suggested that spatial information verification methods discussed in \ref{Spacial} need to be adopted together with the above methods.

\section{Conclusions}
In this paper, we introduce the bag-of-words (BoW) model in image retrieval task, which relies on the local descriptors such as SIFT.
The main idea of BoW is quantizing local descriptors into visual words which form a vocabulary, and then applying scalable textual indexing and search technique.
We then sequentially discuss the key point detection, local description, vocabulary generation, vector quantization, indexing and search.
To further improve the performance of BoW, we also introduce the idea of incorporating spatial information, combining BoW model with global features and query expansion.

\bibliographystyle{plain}
\bibliography{CBIR_BoW}

\begin{thebibliography}{10}

\bibitem{surf}
Herbert Bay, Tinne Tuytelaars, and Luc J.~Van Gool.
\newblock Surf: Speeded up robust features.
\newblock In {\em ECCV (1)}, pages 404--417, 2006.

\bibitem{spatial}
Yang Cao, Changhu Wang, Zhiwei Li, Liqing Zhang, and Lei Zhang.
\newblock Spatial-bag-of-features.
\newblock In {\em CVPR}, pages 3352--3359, 2010.

\bibitem{similarityestimation}
Moses Charikar.
\newblock Similarity estimation techniques from rounding algorithms.
\newblock In {\em STOC}, pages 380--388, 2002.

\bibitem{totalrecall}
Ondrej Chum, James Philbin, Josef Sivic, Michael Isard, and Andrew Zisserman.
\newblock Total recall: Automatic query expansion with a generative feature
  model for object retrieval.
\newblock In {\em ICCV}, pages 1--8, 2007.

\bibitem{fire}
Tobias Gass, Tobias Weyand, Thomas Deselaers, and Hermann Ney.
\newblock Fire in imageclef 2007: Support vector machines and logistic models
  to fuse image descriptors for photo retrieval.
\newblock In {\em CLEF}, pages 492--499, 2007.

\bibitem{similarityhashing}
Aristides Gionis, Piotr Indyk, and Rajeev Motwani.
\newblock Similarity search in high dimensions via hashing.
\newblock In {\em VLDB}, pages 518--529, 1999.

\bibitem{packing}
Herv{\'e} J{\'e}gou, Matthijs Douze, and Cordelia Schmid.
\newblock Packing bag-of-features.
\newblock In {\em ICCV}, pages 2357--2364, 2009.

\bibitem{improving}
HHerv{\'e} J{\'e}gou, Matthijs Douze, and Cordelia Schmid.
\newblock Improving bag-of-features for large scale image search.
\newblock {\em International Journal of Computer Vision}, 87(3):316--336, 2010.

\bibitem{towards}
Yu-Gang Jiang, Chong-Wah Ngo, and Jun Yang.
\newblock Towards optimal bag-of-features for object categorization and
  semantic video retrieval.
\newblock In {\em CIVR}, pages 494--501, 2007.

\bibitem{beyond}
Svetlana Lazebnik, Cordelia Schmid, and Jean Ponce.
\newblock Beyond bags of features: Spatial pyramid matching for recognizing
  natural scene categories.
\newblock In {\em CVPR (2)}, pages 2169--2178, 2006.

\bibitem{sift}
David~G. Lowe.
\newblock Object recognition from local scale-invariant features.
\newblock In {\em ICCV}, pages 1150--1157, 1999.

\bibitem{distinctive}
David~G. Lowe.
\newblock Distinctive image features from scale-invariant keypoints.
\newblock {\em International Journal of Computer Vision}, 60(2):91--110, 2004.

\bibitem{ann}
Marius Muja and David~G. Lowe.
\newblock Fast approximate nearest neighbors with automatic algorithm
  configuration.
\newblock In {\em VISAPP (1)}, pages 331--340, 2009.

\bibitem{hkm}
David Nist{\'e}r and Henrik Stew{\'e}nius.
\newblock Scalable recognition with a vocabulary tree.
\newblock In {\em CVPR (2)}, pages 2161--2168, 2006.

\bibitem{sampling}
Eric Nowak, Fr{\'e}d{\'e}ric Jurie, and Bill Triggs.
\newblock Sampling strategies for bag-of-features image classification.
\newblock In {\em ECCV (4)}, pages 490--503, 2006.

\bibitem{lbp}
Timo Ojala, Matti Pietik{\"a}inen, and David Harwood.
\newblock A comparative study of texture measures with classification based on
  featured distributions.
\newblock {\em Pattern Recognition}, 29(1):51--59, 1996.

\bibitem{object}
James Philbin, Ondrej Chum, Michael Isard, Josef Sivic, and Andrew Zisserman.
\newblock Object retrieval with large vocabularies and fast spatial matching.
\newblock In {\em CVPR}, 2007.

\bibitem{surveyobjectrecognition}
Peter~M. Roth and Martin Winter.
\newblock {Survey of Appearance-Based Methods for Object Recognition}.
\newblock Technical report, Institute for Computer Graphics and Vision, Graz
  University of Technology, 2008.

\bibitem{semantichashing}
Ruslan Salakhutdinov and Geoffrey~E. Hinton.
\newblock Semantic hashing.
\newblock {\em Int. J. Approx. Reasoning}, 50(7):969--978, 2009.

\bibitem{videogoogle}
Josef Sivic and Andrew Zisserman.
\newblock Video google: Efficient visual search of videos.
\newblock In {\em Toward Category-Level Object Recognition}, pages 127--144,
  2006.

\bibitem{spectralhashing}
Yair Weiss, Antonio Torralba, and Robert Fergus.
\newblock Spectral hashing.
\newblock In {\em NIPS}, pages 1753--1760, 2008.

\bibitem{bundling}
Zhong Wu, Qifa Ke, Michael Isard, and Jian Sun.
\newblock Bundling features for large scale partial-duplicate web image search.
\newblock In {\em CVPR}, pages 25--32, 2009.

\end{thebibliography}

\end{document}